\documentclass[aps,prl,twocolumn,showpacs]{revtex4-1}

\usepackage{graphicx} \usepackage{amsmath}
\usepackage[colorlinks]{hyperref}

\newcommand{\EqLabel}[1]{\label{#1}}

\begin{document}
 
\title{Magnon-mediated interactions between fermions depend strongly
  on the lattice structure 
}

\author{Mirko M\"oller, George A. Sawatzky and Mona Berciu} 

\affiliation{Department of Physics \&  Astronomy, University of
  British Columbia, Vancouver, BC, Canada,  V6T 1Z1}

\date{\today}
 
\begin{abstract} 
We propose two new methods to calculate exactly the spectrum of two
spin-${1\over 2}$ charge carriers moving in a ferromagnetic background, at zero
temperature. We find that if the spins are located on a different
sublattice than that on which the fermions move, magnon-mediated
effective interactions are very strong and can bind the fermions into
low-energy bipolarons with triplet character. This never happens in
models where spins and charge carriers share the same lattice, whether they
are in the same band or in different bands. This proves that effective
one-lattice models do not describe correctly the low-energy part of
the two-carrier spectrum of a two-sublattice model, even though they
may describe the low-energy single-carrier spectrum appropriately.
\end{abstract}


\maketitle 

When studying properties of complex materials with ions on several
sublattices it is customary to use simplified, one-lattice
Hamiltonians to describe their low-energy physics.  For example,
instead of a two-sublattice model for the CuO$_2$ plane including both
Cu $3d_{x^2-y^2}$ and O $2p$ orbitals \cite{Emery}, many prefer a
one-band Hubbard model on the Cu (sub)lattice. States in this simpler
model are Zhang-Rice singlets (ZRS), {\em i.e.} bound singlets between
a hole at a Cu site and a doping hole delocalized over its four O
neighbours with the same $d_{x^2-y^2}$ symmetry \cite{ZR}.

Such a composite object may describe well the low-energy
quasiparticle, although this is still debated \cite{Bayo}. Less clear
is whether a model based on such states that mix together charge and
spin degrees of freedom, can properly describe quasiparticle
interactions, especially those mediated through spin 
fluctuations. Most oxides have at least one phase with long-range
magnetic order, and magnon exchange is believed by some to be a key
component determining their properties, {\em eg.} as the main
``glue'' for pairing in cuprates, which
likely controls the value of $T_c$ \cite{Tcrev}.

Here we show that effective one-lattice models severely underestimate
the magnon-mediated attraction between carriers, compared to their
two-sublattice ``parent'' model. The magnetic background is chosen 
as ferromagnetic (FM). This is  much simpler  than an
antiferromagnetic (AFM) background, but it allows for exact
solutions. Thus, any qualitative differences are 
inherent to the models themselves. Moreover, our conclusions are
relevant to the modeling of carriers in AFM backgrounds, and raise
serious questions about the ability of ZRS-like constructs to describe
correctly low-energy two-carrier states.

\begin{figure}[b]
\includegraphics[width=\columnwidth]{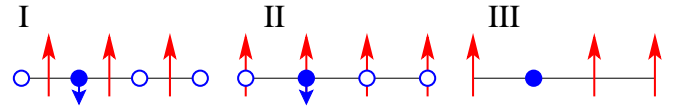}
\caption{Models I and II have two bands: one occupied
  by spins (arrows), and one (empty circles) hosting  carriers
  introduced by doping (filled circles, with arrow showing the spin).
  In the ``parent'' model I, these are on different sublattices. In
  model II, they are on the same lattice. Model III has one band which
  hosts both spins (arrows) and ZRS-like polaron cores (filled
  circle).
\label{fig1}}
\end{figure}

Our models are sketched in Fig. \ref{fig1}. For simplicity, we address
the one-dimensional (1D) case; generalizations are straightforward.
Model I is the ``parent'', two-sublattice, two-band model. Model II is
a two-band, single lattice effective model. Model III is an even
simpler one-band effective model.

In models  I and II, one band hosts the spin $S$ degrees of
freedom, described by:
\begin{equation}
\EqLabel{fm}
\nonumber
{\cal H}_{s} =-J \sum_{i}^{} \left(\vec{S}_i \cdot \vec{S}_{i+1}-S^2\right).
\end{equation}
This favours a FM ground-state $|FM \rangle =|+S,\dots,+S\rangle$ for
the undoped system. Spin-${1\over 2}$ doping charge carriers occupy
states in another band. In model I, this is located on a different
sublattice, for example like in a CuO chain with spins on Cu and holes
on O sites. In model II, they are on the same lattice.
In both cases carriers are described by a Hubbard model:
\begin{equation}
\EqLabel{cc}
\nonumber
{\cal H}_{c} = -t \sum_{i,\sigma}^{}\left(
c^\dagger_{i+1+\delta,\sigma}c_{i+\delta,\sigma} 
+ h.c.\right) + U \sum_{i}^{} \hat{n}_{i+\delta, \uparrow}
\hat{n}_{i+\delta, \downarrow}  
\end{equation}
where $\delta={1\over 2}$ for model I, $\delta=0$ for model
II, $c_{i+\delta,\sigma}$ is the annihilation operator for a
 carrier of spin $\sigma$ from site $i+\delta$,
and $\hat{n}_{i+\delta,\sigma} =
c^\dagger_{i+\delta,\sigma}c_{i+\delta,\sigma}$.

Interactions between charges and spins are
described by the simplest exchange model. In model I, a
carrier interacts with its two  neighbour spins:
\begin{equation}
\EqLabel{exI}
\nonumber
{\cal H}^{(I)}_{ex} = J_0 \sum_{i}^{} \vec{s}_{i+{1\over 2}}\left(\vec{S}_i +
\vec{S}_{i+1}\right), 
\end{equation}
while for model II there is on-site exchange:
\begin{equation}
\EqLabel{exII}
\nonumber
{\cal H}_{ex}^{(II)} = J_0 \sum_{i}^{} \vec{s}_i  \cdot \vec{S}_i.
\end{equation}
Here $\vec{s}_{i+\delta} = {1\over 2}
\sum_{\alpha,\beta}^{}c^\dagger_{i+\delta,\alpha}\vec{\sigma}_{\alpha\beta}
c_{i+\delta,\beta}$, where $\vec{\sigma}$ are Pauli matrices. We set
 $\hbar=1$ and the lattice constant $a=1$.

The one-band model III is described by the Hubbard Hamiltonian ${\cal H}_c$
projected onto the appropriate  subspace, {\em eg.}
$S^z_{tot}=NS-{1\over2}$ if one spin-down carrier is added.

We set $S={1\over 2}$ (higher $S$ is discussed elsewhere \cite{next}),
and first review the one-carrier case. If $\sigma=\uparrow$, the $T=0$
problem is trivial in models I and II: spin-flips are impossible,
therefore the eigenstates $ c^\dagger_{k\uparrow}|FM\rangle$ have
energy $E_{k\uparrow} = \epsilon_k + \gamma J_0$, where
$\gamma={1\over 2}\left(\gamma={1\over 4}\right)$ in model I (II), and
$\epsilon_k=-2t\cos(k)$ is the free carrier dispersion. Model III has
a serious problem: it cannot distinguish an undoped system from one
doped with spin-up carriers.

The interesting case is for a carrier injected with spin-down. In
model III, this is trivial: the ``hole'' moves freely with energy
$\epsilon_k$.  The exact solution (in any dimension) for the Green's
function $G_{\downarrow}(k,\omega) = \langle FM| c_{k,\downarrow}
\hat{G}(\omega) c_{k,\downarrow}^\dagger| FM\rangle$ where
$\hat{G}(\omega) = [\omega+i\eta - {\cal H}]^{-1}$ is the resolvent
for  ${\cal H}={\cal H}_s+{\cal H}_c +{\cal
H}_{ex}$, has long been known  for model II \cite{Shastry}. Its
generalization to two-sublattice models was proposed
recently \cite{FM}. The off-diagonal part of ${\cal 
H}_{ex}$  mixes  $c^\dagger_{k\downarrow}|FM\rangle$, of
energy $ E_{k\downarrow}= \epsilon_k - \gamma J_0$, with the continuum
of one-magnon states $ c^\dagger_{k-q\uparrow}S_q^-|FM\rangle$ of
energy $ E_{k-q,\uparrow} +\Omega_q$. Here, $S_q^- = {1\over \sqrt{
N}} \sum_{i}^{} e^{i q R_i}S_i^-$ is the magnon creation operator and
$\Omega_q= 2J \sin^2(q/2)$ is the spin-wave dispersion. Also,
$S_i^-=S_i^x-iS_i^y$ is the spin-lowering operator and $N
\rightarrow \infty$ is the number of sites on either sublattice.

The interesting case is $J_0>0$: hybridization pushes the discrete
state further below the continuum, resulting in a low-energy
infinitely lived quasiparticle (spin-polaron), discussed next. If
$J_0<0$, the low energy states are the incoherent continuum describing
the scattering of the carrier off a free magnon \cite{Shastry}. One
cannot further simplify the description of such states.

Returning to the spin-polaron that is the low-energy quasiparticle for
$J_0>0$, its structure can be understood in the limit $J_0 \gg
t,J$. We start with model II. ${\cal H}_{ex}^{(II)}$ is minimized by
an on-site singlet between the carrier and its spin, $
|s\rangle_i = {1\over \sqrt{2}} \left( c^\dagger_{i\downarrow}
-c^\dagger_{i\uparrow}S_i^- \right) |FM\rangle,$ with all other spins
in the FM state. The energy of this  degenerate state is ${\cal
H}_{ex}^{(II)} |s\rangle_i = -{3\over 4}J_0 |s\rangle_i$. Hopping
lifts degeneracy, and to first order in $t,J$, the polaron
energy is:
\begin{equation}
\EqLabel{epII}
E_{P}^{(II)}(k) \approx -{3\over 4}J_0 + {1\over 2} \epsilon_k
+ {J\over 2}
\end{equation}
with $ |P_{{\rm II}},k\rangle = {1\over \sqrt{N}} \sum_{i}^{} e^{i k
  R_i} |s\rangle_i$.  Thus, the spin-polaron is an on-site singlet
  between the charge and its local spin (or a bound state of the
  carrier with a magnon at the same site) that moves with an effective
  hopping $t/2$ suppressed by the magnon cloud overlap. The last term
  is the FM exchange energy lost in the magnon's presence.

Similar considerations apply to model I. Again, we only discuss the
case $J_0>0$ which has a low-energy quasiparticle. Because of the
two-sublattice structure, the ground state of ${H}_{ex}^{(I)}$ is the
 three-spin polaron (3sp)
$|3sp\rangle_{i+{1\over 2}} =  \left({\sqrt{2\over 3}}
 c_{i+{1\over 2},\downarrow}^\dagger - c_{i+{1\over
    2},\uparrow}^\dagger \frac{S_i^-+S_{i+1}^-}{\sqrt{6}} \right)|FM\rangle,
$
of energy  ${\cal 
  H}_{ex}^{(I)} |3sp\rangle_{i+{1\over2}} = -J_0
|3sp\rangle_{i+{1\over2}}$  \cite{Emery2}. It describes a bound state between
the carrier and a magnon on either  neighbouring spin. Hopping lifts
degeneracy, resulting in
$|P_{{\rm I}},k\rangle ={1\over \sqrt{N}} \sum_{i}^{} e^{i k
  (R_i+{1\over2}) }|3sp\rangle_{i+{1\over 2}}$
with energy:
\begin{equation}
\EqLabel{epI}
E_{P}^{(I)}(k) \approx -J_0  + {5\over 6} \epsilon_k 
+ {J \over 6}.
\end{equation}

\begin{figure}[t]
\includegraphics[width=\columnwidth]{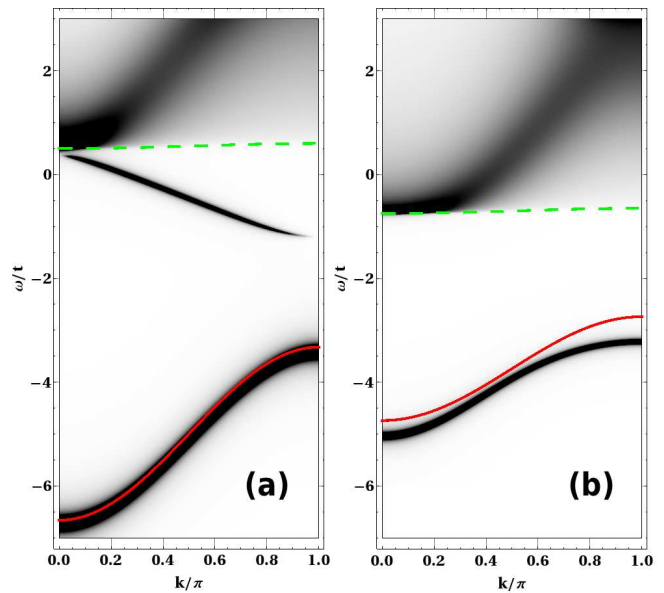}
\caption{(color online) (a) Model I, and (b) model II density of states
  $\rho_{\downarrow}(k,\omega)=-{1\over \pi} \mbox{Im}
  G_{\downarrow}(k,\omega)$. 
  Contour plots show exact results. Full lines are 
  Eqs. (\ref{epII}), (\ref{epI}), while dashed lines mark the
  expected onset of the continuum, at $\min_q \left(E_{k-q,\uparrow} +
  \Omega_q\right)$. Here $J/t=0.05, J_0/t=5,
  \eta/t=0.02$. 
\label{fig2}}
\end{figure}

In Fig. \ref{fig2} we compare the exact spectra, calculated as
 in Refs. \cite{Shastry,FM}, with these asymptotic
expressions (lines) for models I and II. The
agreement is excellent for large $J_0$ and remains reasonable even
down to $J_0\sim t$, showing that this singlet/3sp
description of the spin-polaron is robust. Fig. \ref{fig2}(a)
also shows another quasiparticle below the continuum. This is
based on a higher eigenstate of ${H}_{ex}^{(I)}$ (for more details,
see Ref. \cite{FM}. Here we focus only on the
low-energy physics).

Based on this, it is clear that if we are only interested in
the low-energy spin-down quasiparticle, we can map
model I onto   II and III, after some rescaling of
parameters. The mapping from model I to II is obvious if we rewrite
$
|P_{{\rm I}},k\rangle = {1\over \sqrt{N}} \sum_{i}^{} e^{i k
  R_i}{1\over \sqrt{2}}\left[d_{k,i,\downarrow}^\dagger - d^\dagger_{k,i,\uparrow}
  S_i^- \right]|FM\rangle, 
$
{\em i.e.} a singlet-type state like $|P_{{\rm
    II}},k\rangle $, except that  the carrier is in the
``on-site'' orbital $d_{k,i,\sigma}^\dagger ={1\over \sqrt{3}} \left(e^{i{k\over 2}}
c^\dagger_{i+{1\over 2},\sigma} + e^{-i{k\over 2}}c^\dagger_{i-{1\over
	2},\sigma} \right).$

Thus, $|P_{{\rm I}},k\rangle$ is the direct analog of the ZRS: in the
$k=0$ ground-state, the carrier occupies a linear combination of
neighbour orbitals with the same $s$-type symmetry as the orbital
giving rise to the $S_i$ spin, with which it locks in a singlet.  (If
the orbitals had other symmetries, the hopping matrices would be
different and the phases would change accordingly \cite{FM}). Like in
the ZRS, $d_{k,i,\sigma}^\dagger$ is not orthogonal to its neighbour
$d_{k,i\pm 1,\sigma}^\dagger$. Unlike the ZRS, these ``on-site''
states depend on $k$. This ensures the normalization of $|P_{{\rm
I}},k\rangle$ in the entire Brillouin zone, unlike for the ZRS whose
normalization diverges at $k=0$ \cite{ZR}.

Nevertheless, if we ignore such complications and replace
$d_{k,i,\sigma} \rightarrow c_{i\sigma}$, model I maps onto model II
as far as the low-energy quasiparticle is concerned. Mapping to model
III is the next step of replacing the ZRS with a ``hole'' which lives
in the same band as the spins of the undoped system. Since model III
has a quasiparticle band in the $S^z_{tot}=(N-1){1\over2}$ sector, it
can also be mapped onto the quasiparticle band of model I, after
rescaling.

The important question is whether this mapping between  low-energy
sectors carries on to cases with more carriers.  We  consider the
two-carriers case, and show that model I has low-energy states which
have no equivalent in models II and III. Specifically, magnon-mediated
interactions can stabilize a low-energy bound pair in model I, if
$J_0$ is sufficiently large.  Such states never occur at low energies
in model II, and are impossible in model III. This proves that
modeling the proper lattice structure is essential  to
correctly describe magnon-mediated  interactions.

The two-particle problem also has several cases. The trivial one is
for two spin-up carriers. In models I and II the carriers behave like
non-interacting particles, with eigenstates $c_{k\uparrow}^\dagger
c_{k'\uparrow}^\dagger|FM\rangle$ of energy $E_{k,\uparrow} +
E_{k',\uparrow}$. Model III does not distinguish this case from an undoped
system.

The most interesting case is when one carrier is injected with spin up
and the other with spin down. We analyze it in detail for models
I and II, and then briefly discuss the case where both carriers are
injected with spin-down. We solve the problem exactly  using two
new  methods.

The first method calculates two-particle Green's functions in momentum
space. Because of translational invariance,  total momentum is
conserved so  non-vanishing matrix elements are $ G(k,q,q',\omega) =
\langle k, q | 
\hat{G}(\omega) |k, q'\rangle$, 
where $ |k,q\rangle = c_{{k\over 2} + q, \uparrow}^\dagger c_{{k\over
2} - q, \downarrow}^\dagger |FM\rangle $ is a two-particle state with
total momentum $k$.  Momentum $q$ is not conserved: on-site scattering
changes it, as do magnon-mediated interactions. In the latter, the
spin-down carrier flips its spin creating a magnon, followed by
absorption of the magnon by the other (initially spin-up) carrier. The
magnon's momentum is, thus, transferred from one carrier to the other,
and their spins are exchanged. It is precisely this effective
interaction that interests us.

Because only one-magnon states are accessible from the original state,
we obtain a closed system of equations of motion for these Green's
functions. This is rather similar to the single-carrier case
\cite{Shastry,FM}, however the solution is now less trivial. In the
Supplemental Material \cite{supp} we present 
the steps to reduce it to a closed equation that can be
efficiently solved for a finite chain with $N\sim
100$. However, this becomes costly in higher dimensions, and
may not generalize to other cases ({\em eg.}, more
carriers).

We also propose a real-space solution for the infinite chain which
allows us to find 
the symmetry of the pair (singlet vs. triplet), and also
generalizes to more carriers. It is 
based on the few-particle solution of Ref. \cite{Mona}. Our case is more
complex because when present, the magnon also counts as a
``particle''. Thus, the system switches between having  the
carriers in  states like
$
|k, n\rangle = {1\over \sqrt{N}}\sum_{i}^{} e^{ik(R_i+ \delta + {n\over
	  2})} c^\dagger_{i+\delta, \uparrow} c^\dagger_{i+\delta+n,
	\downarrow}  |FM\rangle 
$
and three ``particle'' configurations
$
|k,n,m\rangle ={ 1\over \sqrt{N}}\sum_{i}^{} e^{ik(R_i+ \delta + {n\over
	  2})} c^\dagger_{i+\delta, \uparrow} c^\dagger_{i+\delta+n,
	\uparrow} S_{i+m}^- |FM\rangle. $
Exchange connects propagators
 $
G(k,n,n',\omega) = \langle k, n| \hat{G}(\omega) | k, n'\rangle$ 
to $G(k,n,n',m,\omega) = \langle k, n| \hat{G}(\omega) |
k, n',m\rangle$, and vice-versa. The solution of these equations is
described in \cite{supp}. 

\begin{figure}[t]
\includegraphics[width=\columnwidth]{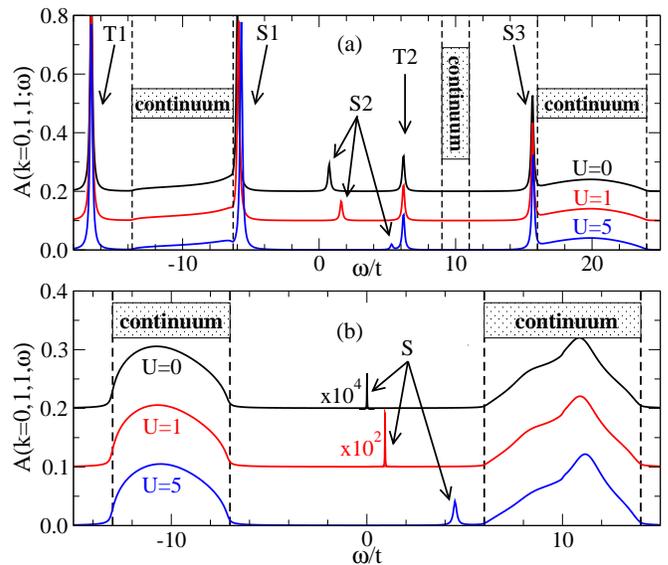}
\caption{(color online) Spectral weight $A(k=0,n=n'=1,\omega)$ for
  model I (a) and II (b). Expected continua locations are
  marked, as are triplet (T) and singlet (S) bipolarons (arrows). Here
  $J_0/t=20, J/t=0.05, \eta/t=0.1$ and 
  $U/t=0,1,5$. 
\label{fig3}}
\end{figure}

Poles of these propagators mark the two-carrier
spectrum. Fig. \ref{fig3} shows $A(k,n, n', 
\omega) = -{1\over \pi} \mbox{Im} 
G(k,n,n',\omega)$ for $k=0, n=n'=1$
 and $J_0=20t$, chosen so large  to simplify
the task of identifying features.

The expected features in the spectrum of model I are: (i) a continuum
describing states where the spin-down carrier forms a polaron of
momentum $k'$ and the spin-up carrier has momentum $k-k'$. These
states span $\{E_P^{(I)}(k') + E_{k-k',\uparrow}\}_{k'}$ so the
band-edges are known from single-carrier results. Using
Eq. (\ref{epI}) and ignoring small $J$ corrections, the band edges are
here expected at $-13.7t$ and $-6.3t$ (dashed lines), in good
agreement with the results; (ii) a continuum $\left\{E_{k',\uparrow} +
E_{k-k'-q,\uparrow} + \Omega_q\right\}_{k',q}$ describing states
comprised of two spin-up carriers and a free magnon. Expected
band-edges at $16t$ and $24t$ (dashed lines) show again  good
agreement; (iii) if there is a higher energy polaron state, like in
Fig. \ref{fig2}(a), it will also generate a continuum. Here, its
edges should be at $9t$ and $11t$. It is not seen in
Fig. \ref{fig3}(a) because of vanishing overlap with the
$|k=0,n=1\rangle$ state, but it appears in momentum-space Green's
functions \cite{supp}.

Any states outside these continua are bound states, ``glued'' through
magnon-exchange. Five such bipolarons appear in Fig. \ref{fig3}(a)
(the apparent overlap between S1 and S3 with their nearby continua is
due to the broadening $\eta=0.1t$).  For $k=0$, the real-space
solution allows us to identify two as triplets and the other as
singlets (see arrows). For finite $k$, there is no definite symmetry:
these bound states have finite overlap with both singlet and triplet
configurations, such as $\sum_{i}^{}
e^{ikR_i}\left(c^\dagger_{i+{1\over 2}, \uparrow} c^\dagger_{i+{1\over
    2}+n, \downarrow}\mp c^\dagger_{i+{1\over2}, \downarrow}
c^\dagger_{i+{1\over 2}+n, \uparrow} \right) |FM\rangle.$

\begin{figure}[t]
\includegraphics[width=\columnwidth]{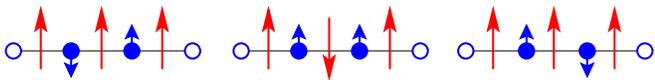}
\caption{(color online) The three highest weight configurations
  contributing to  the low-energy
  bipolaron of model I. 
\label{fig4}}
\end{figure}

Consider now the low-energy bipolaron. Its largest overlap is with the
configurations sketched in Fig. \ref{fig4}, hybridized through the
$J_0$ exchange. It is the ability of both carriers to  interact with
{\em the same spin} to exchange the magnon, that stabilizes this state. This
also explains why it is a triplet at $k=0$ (symmetric combination is
favoured), and its insensitivity to $U$: unlike singlets, triplets do
not permit double occupancy. This bipolaron forms if $J_0/t>
6.5$, and is quite light \cite{next}.

Model II has different behaviour, see Fig. \ref{fig3}(b). Again, a
polaron+spin-up carrier continuum, and a magnon + two spin-up carriers
continuum, are expected and observed. If $J_0$ is sufficiently large
that the two do not overlap, a bound state appears between them, at
$\omega\sim U$. This bipolaron is a singlet at $k=0$, quite similar to
the S2 state in model I. It has very little weight in the
$|k=0,n=1\rangle$ configuration, which is why we had to magnify the
peaks. Most of its weight is on the on-site $(n=0)$ configuration
\cite{supp}.  No triplet-like bound states appear, and no
low-energy bipolaron is possible for any values of the
parameters. This is not so surprising, since in this model
only one  carrier can interact with a given spin/magnon. If the
carriers are on the same site they form a singlet which has no
interactions with the local spin. If they are on neighbouring sites,
 magnon exchange (now controlled by $J$, not by
$J_0$ as in Fig. \ref{fig4}) is too weak to stabilize a low-energy
bipolaron. Thus, model II simply cannot describe the low-energy
physics of model I in this sector. Model III fails as well, since it
does not distinguish between a spin-up carrier and a lattice spin.

We also considered the two spin-down carrier states. For models I and
II, the $t=0$ solutions suggest that the lowest energy feature is the
two-polaron continuum. This is reasonable, since each carrier can bind
its own magnon to create a polaron. Simultaneous interaction of one
carrier with two magnons is impossible in model II, and while
possible, it is energetically unfavourable in model I, so low-energy
bipolarons do not appear in this sector. Turning on hopping further
favours the continuum, since polarons are lighter than a bipolaron
\cite{next}.  In model III, no interactions are possible between two
``holes'', since they simply reshuffle the FM spins as they move. This
is why we have not investigated this case in more detail.

To summarize, we extended the exact solution for a single charge in a
FM background to cases with two or more carriers, and discussed in
detail the nontrivial case where one carrier is injected with spin up
and the other with spin down. The low-energy physics depends
essentially on the model. If the spins are intercalated between the
carrier sites, magnon-exchange is enhanced and can bind low-energy
bipolarons. If spins and carriers live on the same lattice, such
low-energy states are impossible. If spins and carriers live in the
same band, there are no magnon-exchange interactions in any of the
allowed cases.  These three models have very different low-energy
states in the two-particle sector, even though their one-particle
sectors can be mapped onto one another.

This shows that in order to properly describe magnon-mediated
interactions, one must use the proper sublattice structure, not
simpler effective one-lattice models. While our work is for a FM
background, it is directly relevant for AFM backgrounds as well, since
the magnon exchange is a rather local process. Based on our results,
for a CuO$_2$ lattice one should expect strong magnon-mediated
interactions between two holes located at neighbour O sites, through
their common Cu (the 2D analog of Fig. \ref{fig4}). A one-band model based on 
ZRSs simply cannot describe this process, and is therefore
likely to severely underestimate the role of magnons as the ``glue''
for pairing.

{\em Acknowledgements:} this work was supported by NSERC, QMI and  CIfAR.


\end{document}